\documentclass[10pt,a4paper,dvipdfmx]{article}
\usepackage{aer}
\usepackage{aer-osborn}
\usepackage{graphicx}
\usepackage[small,bf,hang,margin=7truemm]{caption}[2007/04/09]
\usepackage{amsmath}
\usepackage[]{natbib}
\usepackage{url}
\usepackage{bm}

\newcommand\eref[1]{Eq.(\ref{#1})}
\newcommand\tabref[1]{Table~\ref{#1}}
\newcommand\figref[1]{Fig.~\ref{#1}}
\newcommand\secref[1]{Section~\ref{#1}}

\newcommand\vek[1]{\bm{#1}}
\newcommand\mat[1]{\mathbf{#1}}
\newcommand\T{{\mathsf{T}}}  

\begin{document}

\def\addnum#1{$^#1$}

\title{\Large\bf
  Structure and temporal change of the credit network between banks and
  large firms in Japan}

\author{\sl 
  Yoshi Fujiwara   \addnum1,
  Hideaki Aoyama   \addnum2,
  Yuichi Ikeda     \addnum3,\\\sl
  Hiroshi Iyetomi  \addnum4,
  Wataru Souma     \addnum1
  \\[10pt]
  {\sl\small\addnum1
    ATR/NiCT CIS Applied Network Science Laboratory, Kyoto 619-0288, Japan}\\
  {\sl\small\addnum2
    Department of Physics, Kyoto University, Kyoto 606-8501, Japan}\\
  {\sl\small\addnum3
    Hitachi Ltd, Hitachi Research Laboratory, Ibaraki 319-1221, Japan}\\
  {\sl\small\addnum4
    Department of Physics, Niigata University, Ikarashi, Niigata 950-2181, Japan}}

\date{}
\maketitle

\begin{abstract}
  We present a new approach to understanding credit relationships
  between commercial banks and quoted firms, and with this approach
  examine the temporal change in the structure of the Japanese credit
  network from 1980 to 2005. At each year, the credit network is regarded as a weighted
  bipartite graph where edges correspond to the relationships and
  weights refer to the amounts of loans.  Reduction in the supply of
  credit affects firms as debtor, and failure of a firm influences
  banks as creditor. To quantify the dependency and influence between
  banks and firms, we propose a set of scores of banks and firms, which can be
  calculated by solving an eigenvalue problem determined by the weight of
  the credit network. We found that a few largest eigenvalues and
  corresponding eigenvectors are significant by using a null
  hypothesis of random bipartite graphs, and that the scores can quantitatively
  describe the stability or fragility of the credit network during the
  25 years.
\end{abstract}



\noindent
{\bf Correspondence\/}:
  Yoshi Fujiwara,
  ATR CIS Applied Network Science Lab,\\
  Kyoto 619-0288, Japan.
  Email: yoshi.fujiwara@gmail.com\\[20pt]

\noindent\hrulefill

\noindent
{\small\sl We would like to thank M.~Gallegati and G.~De~Masi for discussion
during a preliminary stage of this work. Y.~F. thank them for the
collaboration \citep{masi2008ajc}. We acknowledge the Nikkei Media
Marketing, Inc.~for technical assistance.}

\newpage
\section{Introduction}\label{sec:intro}

The credit-debt relation between banks and firms is one of the most
important relationships among economic agents. Credit is a source of profit
for a bank, and it is fuel for growth of a firm. The flip side
of the relation is, however, the path where failures take place and
their propagation occurs often at a nation-wide scale, and sometimes to
a world-wide extent, as we experience today.

It is well known that the Japanese banking system suffered a
considerable deterioration in its financial condition during the
1990s. Financial institutions in private-sector had accumulated loan
losses, more than 80 trillion yen (nearly 15\% of GDP), which reduced
the bank capitalization, and led to the failure of three major and
other small banks. Even though two major banks were nationalized in
1997, and other political decisions were made in order to maintain the
stability of financial system, most banks, major and minor, decreased
the supply of credit immediately; even by reducing existing loans to
firms. A lot of firms, especially small and medium-sized firms,
eventually suffered loss of funding. See \citet{brewer2003brd}.

Financial systems are, at an aggregate level, subject to the tails of
distributions for economic variables. This perspective has been
recognized increasingly in economics; personal income, firm-size,
number of relationships among firms and banks (ownership,
supplier-customer, etc.), and so on. It has been recognized that
distributions and fluctuations are the keys for understanding many
phenomena in macro-economy (see \citet{aoki2007rme} and
\citet{delligatti2008eme}).

\begin{figure}[htb]
  \centering
  \includegraphics[width=0.70\textwidth]{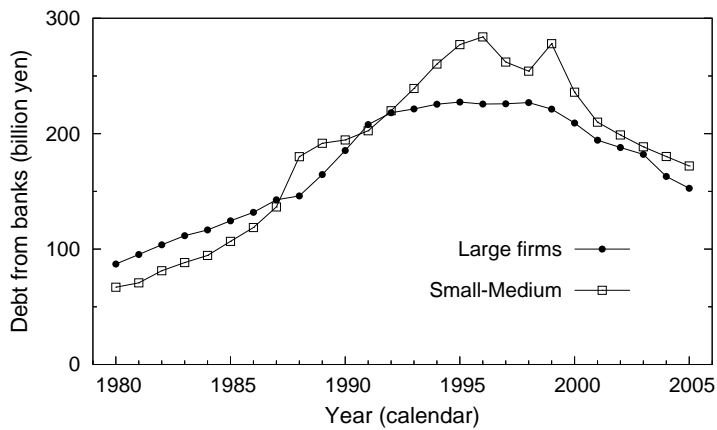}
  \caption{%
    Historical data of the total amount of debt from banks during the
    calendar years, 1980 to 2005. For large firms (filled circles) and
    for small and medium firms (squares).}
\label{fig:doe_sml}
\end{figure}

\figref{fig:doe_sml} shows the historical data of the total amount of
debt from banks for large firms and for small and medium firms%
\footnote{%
  Source: 2008 white papers on small and medium enterprises in Japan,
  Small and Medium Enterprise Agency. Here large firms are the
  companies capitalized at 100 million yen or more, and small-medium
  firms are the others. Calendar years are used here and throughout
  this paper.}%
. For the year 2005, 1.25\% (33,833) of domestic firms are the large
firms according to the classification, while the rest 98.75\% are the
small-medium firms%
\footnote{%
  Source: statistics of corporations by industry, annual report, 1980
  to 2005, Ministry of Finance.}%
. Yet the total loans for the large firms amount to be 160
billion yen, which is nearly equal to those for the small-medium
firms as shown in the figure. Thus, only a small fraction of firms account for half of
all loans. Conversely, as we shall show in this paper, a large part
of loans is provided by a few large banks --- the tail of another
distribution.

Suppose a large firm is heavily indebted with banks. Then a failure of
the firm, or a default, may cause a considerable effect on the balance
sheets of the banks. If the banks reduce their supply of credit,
then the total supply of loans will be decreased resulting in the
adverse shocks to other firms. Therefore, the study of structure of
credit relationships or {\it credit network\/} between banks and
firms, and its temporal change would give us an insight to understand
the financial stability or fragility. This is precisely the purpose of
this paper.

There are several related works in the literature. For example,
\citet{ogawa2007jfp} carried out an analysis of dependency of
the number of long-term credit relationships on characteristics of
firms. \citet{uchida2008bsl} studied the relation between bank-size
and credit links. \citet{kano2006ivb} investigated the credit of small
and medium-sized firms. Studies such as \citet{ogawa2007jfp} focus on
multiple lending relationships. Recently, complex
network analysis (see \citet{caldarelli2007sfn} and references
therein) has been applied to financial systems (e.g.,
\citet{inaoka2004ssb}, \citet{imakubo2008nft}, \citet{iori2007nai} for
inter-bank relationships, \citet{masi2007bft}, \citet{masi2008ajc} for
bank-firm relationships).
In this paper, we shall study on the credit network between banks and
large firms by regarding the network as a {\it weighted bipartite
  graph}, develop quantification of fragility of banks, and apply it
to credit networks in Japan for the past 25 years.

In \secref{sec:data}, we describe our credit network dataset.  In
\secref{sec:nw_def}, we consider a credit network as a
weighted bipartite graph, and show several statistical properties of
heavy-tailed distributions. Then, in
\secref{sec:nw_frag}, we propose a set of scores for
banks and firms which measure potential influences that one agent
exerts on the other. It is shown that the scores can be calculated by
solving an eigenvalue problem. In \secref{sec:nw_res},
we apply this method to our dataset from the year 1980 to 2005. The
results are discussed in \secref{sec:disc}. \ref{sec:math} is
for proving mathematical properties for the eigenvalue problem which
appeared in \secref{sec:nw_frag}.

\section{Dataset}\label{sec:data}

Our dataset is based on a survey of firms quoted in the Japanese
stock-exchange markets (Tokyo, Osaka, Nagoya, Fukuoka and Sapporo, in
the order of market size). The data were compiled from the firms'
financial statements and survey by {\it Nikkei Media Marketing,
  Inc.} in Tokyo, and are commercially available. They include the
information about each firm's borrowing obtained from financial institutions
such as the amounts of borrowing and their classification into short-term and long-term
borrowings. We examined the period from the years 1980 to 2005, for
which incomplete data are few, and study the time development of credit
relationships by using the total of long and short-term credit.

For financial institutions, we select commercial banks as a set of
leading suppliers of credit. The set comprises long-term, city,
regional (primary and secondary), trust banks, insurance companies and
other institutions including credit associations. During the examined
period, more than 200 commercial banks existed, which are summarized in
\tabref{tab:banks}. We remark that failed banks are included until the
year of failure, and that merger and acquisition of banks are processed
consistently. For quoted firms, we choose only
{\it surviving\/} firms that are quoted in the stock markets mentioned
above%
\footnote{%
  Based on the lists of surviving firms and quoted firms in September
  and December 2007 respectively. Firms registered on over-the-counter
  (OTC) market and/or on JASDAQ (the present OTC market) are excluded.
  The dataset includes the OTC and JASDAQ data since 1996, so we
  exclude them also by checking the listing date of the firms added in
  the dataset.}%
.

\begin{figure}[!tb]
  \centering
  \includegraphics[width=0.70\textwidth]{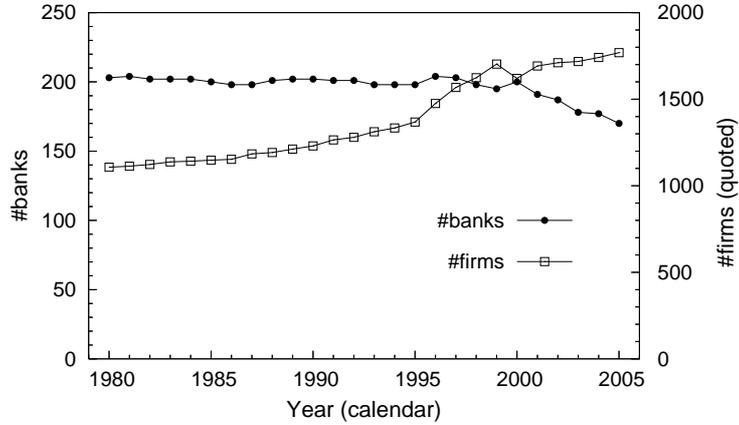}
  \caption{%
    The number of commercial banks and quoted firms.}
\label{fig:ts_numbf}
\end{figure}

The number of banks and firms in each year is summarized in
\figref{fig:ts_numbf}. The classification of banks and industrial
sectors of firms are shown in \tabref{tab:banks} and \tabref{tab:firms}
respectively.

\begin{table}[!tb]
  \centering
  \caption{%
    Classification of commercial banks. \# denotes the
    net number of institutions in each corresponding category during
    the years, 1980 to 2005. The leftmost column, {\sl a\/} to {\sl j\/},
    is defined as a short-hand notation.}
  \label{tab:banks}
  \begin{tabular}{|l|l|r|}
\hline
id & Classification & \# \\
\hline\hline
{\sl a} & Long-term credit banks & 3  \\ 
{\sl b} & City banks & 16  \\ 
{\sl c} & Regional banks & 64  \\ 
{\sl d} & Secondary regional banks & 71  \\ 
{\sl e} & Trust banks & 20  \\ 
{\sl f} & Life insurance companies & 23  \\ 
{\sl g} & Non-life insurance companies & 23  \\ 
{\sl h} & Credit associations (Shinkin banks) & 4  \\ 
{\sl i} & Agricultural financial institutions & 4  \\ 
{\sl j} & Shoko Chukin bank & 1  \\ 
\hline
& Total & 229  \\
\hline
  \end{tabular}
\end{table}

The number of banks and firms in each year is summarized in
\figref{fig:ts_numbf}. The classification of banks and industrial
sectors of firms are shown in \tabref{tab:banks} and \tabref{tab:firms}
respectively.

\begin{table}[!tb]
  \centering
  \caption{%
    Sectors of quoted firms in the dataset. \# denotes the net number
    of firms in each sector during the years, 1980 to 2005. The total
    number of the firms amounts to 2,330.}
  \label{tab:firms}
  \begin{tabular}{|l|r||l|r|}
\hline
manufacturing & \# & non-manufacturing & \# \\
\hline\hline
Foods & 105 & Marine products & 5 \\
Textile products & 60 & Mining & 7 \\
Pulp \& paper & 18 & Construction & 148 \\
Chemicals & 156 & Wholesale trade & 233 \\
Drugs \& medicines & 33 & Retail trade & 153 \\
Petroleum \& coal & 11 & Securities & 18 \\
Rubber products & 20 & Credit \& leasing & 75 \\
Ceramic, etc. & 49 & Real estate & 75 \\
Iron \& steel & 49 & Railway transport. & 27 \\
Non-ferrous metals & 106 & Road transport. & 28 \\
General machinery & 182 & Water transport. & 15 \\
Electronics & 203 & Air transport. & 4 \\
Shipbuilding & 6 & Warehousing & 38 \\
Motor vehicles & 65 & Information Tech. & 20 \\
Transportation equip. & 11 & Utilities (electric) & 11 \\
Precision instruments & 40 & Utilities (gas) & 13 \\
Other manufacturing & 82 & Services & 264 \\
\hline
  \end{tabular}
\end{table}

\section{Analysis of Credit Network}\label{sec:nw}

\subsection{Credit Network as a Weighted Bipartite Graph}\label{sec:nw_def}

\begin{figure}[htbp]
  \centering
  \includegraphics[width=0.40\textwidth]{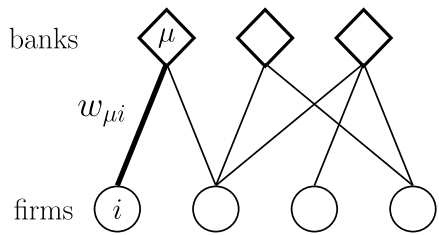}
  \caption{%
    Credit network as a bipartite graph. An edge connecting between
    bank $\mu$ and firm $i$ is associated with an amount of credit
    $w_{\mu i}$ as a weight.}
\label{fig:defw}
\end{figure}

Each yearly statement, or snapshot, of the credit network in our
dataset can be regarded as a bipartite graph. Nodes are either banks
or firms\footnote{%
  Note that banks are not included in the side of firms, even if they
  are borrowing from other banks. Because our dataset includes banks'
  borrowing only partially, the interbank credit is not considered
  here, though it is no less important than the bank-firm credit
  studied here.}%
. Banks and firms are denoted by Greek letters $\mu$
($\mu=1,\ldots,n$) and Latin letters $i$ ($i=1,\ldots,m$)
respectively. $n$ is the number of banks, and $m$ is that of firms.
An edge between a bank $\mu$ and a firm $i$ is defined to be present
if there is a credit relationship between them. In addition, a
positive {\it weight\/} $w_{\mu i}$ is associated with the edge, which
is defined to be the amount of the credit. We can depict the network
as shown in \figref{fig:defw}.

$w_{\mu i}$ is the amount of lending by bank $\mu$ to firm $i$, which
precisely equals to the amount of borrowing by firm $i$ from bank
$\mu$. The total amount of lending by bank $\mu$ is
\begin{equation}
  w_\mu:=\sum_i w_{\mu i}\ ,
\label{eq:sb}
\end{equation}
and the total amount of borrowing by firm $i$ is
\begin{equation}
  w_i:=\sum_\mu w_{\mu i}\ .
\label{eq:sf}
\end{equation}

We note that a same value $w_{\mu i}$ has different meanings as a
weight to the bank $\mu$ and to the firm $i$. For example, even if 90\%
of the total lending of the bank $\mu$ goes to the firm $i$, it may be
the case that $i$ depends on $\mu$ by only 20\% for all the loans from banks.
It would be natural to define an $(n\times
m)$ matrix $\mat{A}$ whose component is given by
\begin{equation}
  A_{\mu i}:=\frac{w_{\mu i}}{w_\mu}\ .
\label{eq:def_A}
\end{equation}
$A_{\mu i}$ represents the relative amount of lending by bank $\mu$ to
firm $i$. We have
\begin{equation}
  \sum_i A_{\mu i}=1\quad\text{for all }\mu\ .
\label{eq:sum_A}
\end{equation}
Similarly, we define an $(m\times n)$ matrix $\mat{B}$ by
\begin{equation}
  B_{i \mu}:=\frac{w_{\mu i}}{w_i}\ .
\label{eq:def_B}
\end{equation}
$B_{i \mu}$ represents the relative amount of borrowing by firm $i$
from bank $\mu$. We have
\begin{equation}
  \sum_\mu B_{i \mu}=1\quad\text{for all }i\ .
\label{eq:sum_B}
\end{equation}

The {\it degree\/} $k_\mu$ of bank $\mu$ is the number of edges emanating
from it to firms, and the degree $k_i$ of firm $i$ is the number of edges
to banks. When the weights $w_{\mu i}$ are all equal to 1, it is
obvious that $k_\mu=w_\mu$ and $k_i=w_i$.

\begin{figure}[tbp]
  \centering
  \includegraphics[width=0.95\textwidth]{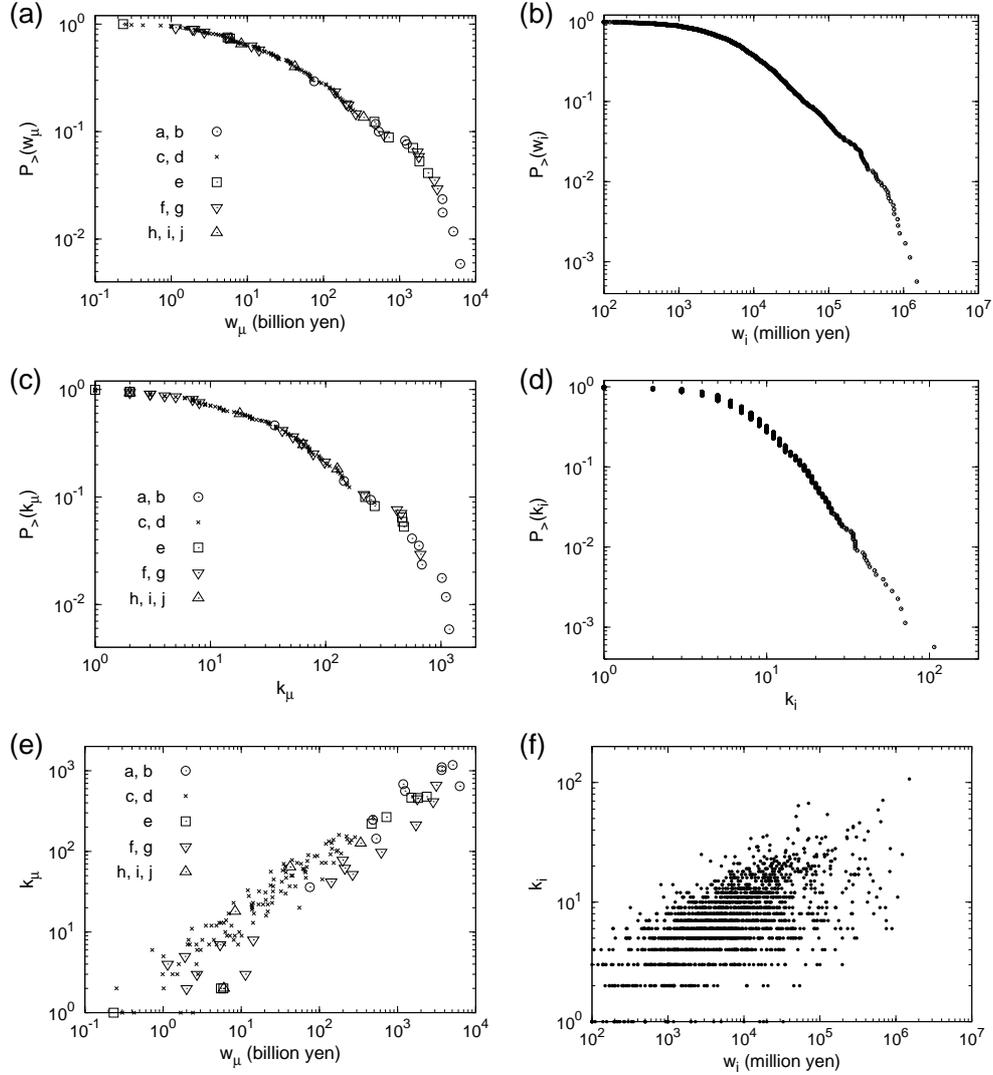}
  \caption{%
    (a) Cumulative distribution $P_>(w_\mu)$ for banks' lending
    $w_\mu$. %
    (b)~$P_>(w_i)$ for firms' borrowing. %
    (c)~$P_>(k_\mu)$ for the number of banks' lending relationships. %
    (d)~$P_>(k_i)$ for the number of firms' borrowing relationships. %
    (e)~Scatter plot for banks' $w_{\mu}$ and $k_{\mu}$. %
    (f)~Scatter plot for firms' $w_i$ and $k_i$. %
    All the plots are for the data in the year 2005. In the plots
    (a),(c) and (e) for banks, the points are drawn according to the
    classification given in \tabref{tab:banks}. %
    Rank correlations (Kendall's $\tau$) for (e) and (f) are
    $\tau=0.825(16.0\sigma)$ and $\tau=0.450(28.3\sigma)$ respectively
    ($\sigma$ calculated under the null hypothesis of statistical
    independence).}
\label{fig:pdfs}
\end{figure}

The distributions for $w_\mu$, $w_i$, $k_\mu$, $k_i$ have long-tails.
They are shown, for the data of credit relationships in the year 2005,
in \figref{fig:pdfs}~(a) to (d). The long-tails for the banks'
amount of credit and number of firms for lending, in
\figref{fig:pdfs}~(a) and (c) for $w_\mu$ and $k_\mu$ respectively,
are comprised of city banks, long-term credit banks, several of
trust banks and insurance companies (see the classification in
\tabref{tab:banks}). Similar long-tails are observed for firms, as shown in
\figref{fig:pdfs}~(b) and (d) for $w_i$ and $k_i$.

There is a significant correlation between $w_\mu$ and $k_\mu$ in a
natural way, and also for $w_i$ and $k_i$, as shown in
\figref{fig:pdfs}~(e) and (f) respectively. We calculated rank
correlation in terms of Kendall's $\tau$, which gave significant
values of $\tau=0.825(16.0\sigma)$ and $\tau=0.450(28.3\sigma)$
respectively, where $\sigma$ is the value under the null hypothesis of
statistical independence. In particular, from the
\figref{fig:pdfs}~(e), we can observe an empirical relation of
$k_\mu\propto w_\mu^a$, where $a\approx0.69\pm0.03$ (least-square fit;
error 95\% level). This implies the relation of $w_\mu/k_\mu\propto
k_\mu^{0.44\pm0.07}$ meaning that the average loan is larger for the
larger degree $k_\mu$, or roughly speaking, for the larger banks. This
observation is consistent with known empirical facts (see
\citet{uchida2008bsl} on similar relation for borrowing by small and
medium-sized enterprises).

We refer the reader to \citet{masi2008ajc} for extensive study on
statistical properties of credit topology and weights.

\subsection{Fragility Scores of Banks}\label{sec:nw_frag}

Bank and firm establish a credit relationship for obvious
reasons. Bank supplies credit in anticipation of interest margin,
and firm uses credit as an important source of financing in
anticipation of growth in its business. An edge of credit, therefore,
represents dependency of one agent on the other in two
ways.

$A_{\mu i}$ quantifies the dependency of bank $\mu$ on firm $i$ as a
source of profit. Also $B_{i \mu}$ is the dependency of firm $i$ on
bank $\mu$ as a source of financing from financial institutions. The
flip side of dependency is a potential influence which one agent exerts
on the other, as we argue below.

\begin{figure}[htbp]
  \centering
  \includegraphics[width=0.75\textwidth]{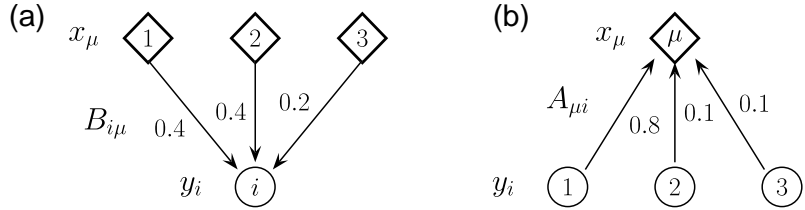}
  \caption{%
    Illustration of (a) influence to firm $i$ from banks with an
    example of weights, $B_{i\mu}$ ($\mu=1,2,3$) satisfying
    \eref{eq:sum_B}, and of (b) influence to bank $\mu$ from firms
    with an example of weights, $A_{\mu i}$ ($i=1,2,3$) satisfying
    \eref{eq:sum_A}.}
\label{fig:inf}
\end{figure}

Suppose that one can quantify a change in the level of bank $\mu$'s
financial deterioration by a variable or score, $x_\mu$, which is to
be defined in a consistent way by the following argument. Bank $\mu$
with increasing $x_\mu$ will behave in various ways; it may shrink the
amount of its supplied credit, increase interest-rate, shorten the due
time of payment by firms, and so forth. In any case, it would
influence firm $i$ to an extent that can be quantified by $B_{i \mu}$,
because it represents the dependency of firm $i$ on bank $\mu$ for the
source of financing. Suppose additionally that a change in the level
of firm $i$'s financial degradation is quantified by another score,
$y_i$, it would be reasonable to assume that $y_i$ is proportional to
$B_{i\mu}\,x_\mu$ summed over banks $\mu$, or $y_i\propto\sum_\mu
B_{i\mu}\,x_\mu$, as the influence from banks to firms.
\figref{fig:inf}~(a) illustrates this direction of influence.

Similarly for the reverse direction of influence, from firms to banks.
Firm $i$ with $y_i$ may delay its repayment, have defaults, even fail
into bankruptcy, and so forth, due to its financial difficulties. Then
the lending banks will not be able to fully enjoy profits in expected
amounts due to the delay, may possibly have bad loans partially, if
not totally, for the credit given to bankrupted firms. Any of them
would result in the banks' financial deterioration, the level of which
was assumed to be quantified by $x_\mu$ at the outset of our argument.
Such influence to bank $\mu$ from firms is reasonably supposed to take
the form, $x_\mu\propto\sum_i A_{\mu i}\,y_i$, in a similar way (see
\figref{fig:inf}~(b) for illustration).

Expressing the change in the level of financial degradation of them as
financial ``fragility'', our consideration above leads us to think
about the influence from one score of fragility to the other by a set
of equations which express the influence:
\begin{align}
  \vek{y} &\propto\mat{B}\vek{x}\ , \label{eq:yx}\\
  \vek{x} &\propto\mat{A}\vek{y}\ , \label{eq:xy}
\end{align}
where $\vek{x}$ and $\vek{y}$ are the vectors with components, $x_\mu$
and $y_i$, respectively. It then follows that
\begin{equation}
  \mat{P}\,\vek{x}=\lambda\,\vek{x}\ ,
\label{eq:eigx}
\end{equation}
where $\mat{P}:=\mat{A}\mat{B}$, $\lambda$ is its eigenvalue and
$\vek{x}$ is the corresponding eigenvector. $\vek{x}$ is the
fragility scores of banks, and $\vek{y}$ is for firms.

Mathematically, an alternative set of scores could be defined, which
we call ``dual'' scores. Namely, they are $u_\mu$ for bank $\mu$ and
$v_i$ or firm $i$ which satisfy
\begin{align}
  \vek{v} &\propto\mat{A}^\T\vek{u}\ , \label{eq:vu}\\
  \vek{u} &\propto\mat{B}^\T\vek{v}\ . \label{eq:uv}
\end{align}
This leads to another eigenvalue problem,
$\mat{P}^\T\,\vek{u}=\lambda\,\vek{u}$, or equivalently
\begin{equation}
  \vek{u}^\T\,\mat{P}=\lambda\,\vek{u}^\T\ .
\label{eq:eigu}
\end{equation}
Here and hereafter, $\T$ represents the transpose of a matrix or a
vector, and we suppose a vector as a column vector by convention.

Thus, the set $\vek{x}$ of fragility scores of banks is the right
eigenvector of the weight matrix $\mat{P}$ as in \eref{eq:eigx}, and
the set $\vek{u}$ of dual scores of banks satisfies the left eigenvector
of $\mat{P}$ as in \eref{eq:eigu}. Since the matrix $\mat{P}$ is not
symmetric, one has a non-trivial relationship between $\vek{x}$ and
$\vek{u}$ due to the definitions of $\mat{P}$ and the weight matrices
of $\mat{A}$ and $\mat{B}$. In \ref{sec:math}, we prove that
the left eigenvalues and the right eigenvalues have a same spectrum,
and that the left eigenvector $\vek{u}$ can be calculated from the
right eigenvector $\vek{x}$ as in \eref{eq:xu}. We shall focus only
on the fragility score in what follows.

As proved in \ref{sec:math}, the eigenvalues and
corresponding eigenvectors have the following mathematical properties.
\begin{itemize}
\item Spectrum of $\lambda$
  \begin{equation}
    0 < \lambda\leq 1\ .
  \label{eq:spectrum}
  \end{equation}
\item Trivial largest eigenvalue:
  \begin{equation}
    \lambda=1\ \text{if and only if}\ x_\mu=\text{constant}.
  \label{eq:trivia}
  \end{equation}
\item Summation formula of the eigenvalues:
  \begin{equation}
    \sum_{k}\lambda_k=\sum_{\mu,i}A_{\mu i}B_{i\mu}
    =\textrm{tr}\,\mat{P}\ .
    \label{eq:lamsum}
  \end{equation}
\end{itemize}

We can also interpret the definition of fragility scores in terms of
dynamical propagation of influence. Let us consider a perturbation, or
an idiosyncratic shock, that occurs with a configuration
$\vek{x}$ among banks. It is assumed that the shock propagates
by \eref{eq:yx} to generate $\vek{y}$ among firms, which in turn
affects the banks by \eref{eq:xy}. Although we do not have knowledge
on the time-scale for this diffusion process, it would be reasonable
to assume that the structure of credit network does not change much in
the meanwhile. Then the propagation of the perturbation, going back
and forth from banks to themselves, could be described by the
repetition of \eref{eq:yx} and \eref{eq:xy}, or equivalently,
$\mat{P}^r$ for a finite number of iterations $r$.

Suppose that the eigenvalues are sorted in the decreasing order:
\begin{equation}
  1=\lambda_1\geq\lambda_2\geq\lambda_3\cdots\lambda_n>0\ .
\end{equation}
The subspace spanned by the trivial eigenvector $\vek{x}^{(1)}$ should
be ignored in the consideration of perturbation, since it merely
represents a constant mode. Denote the resulting vector by
$\tilde{\vek{x}}$, and expand it with respect to the non-trivial
eigenvectors as $\tilde{\vek{x}}=\sum_{k=2}^n a_k\,\vek{x}^{(k)}$,
then
\begin{align}
  \mat{P}^r\tilde{\vek{x}} &=
  \lambda_2^r\,a_2\,\vek{x}^{(2)}+
  \lambda_3^r\,a_3\,\vek{x}^{(3)}+\cdots+
  \lambda_n^r\,a_n\,\vek{x}^{(n)} \nonumber\\
  &= \lambda_2^r\left[a_2\,\vek{x}^{(2)}+
    \biggl(\frac{\lambda_3}{\lambda_2}\biggr)^r a_3\,\vek{x}^{(3)}+\cdots+
    \biggl(\frac{\lambda_n}{\lambda_2}\biggr)^r a_n\,\vek{x}^{(n)}\right]
  \ .
\end{align}
This shows that the behavior of perturbation, in a long run
$r\rightarrow\infty$, is determined mainly by the second largest
eigenvalue and its corresponding eigenvector. For a finite $r$, it is
suggested that one should consider only a few largest eigenvalues and
the corresponding eigenvectors.

Therefore, the eigen decomposition of the idiosyncratic shocks, the
profile of which is not known beforehand, can tell us which
eigen-modes are important in the propagation of influence from banks
to firms and {\it vice versa\/} in a finite time-scale.

\subsection{Results for the Dataset}\label{sec:nw_res}

One needs to evaluate which eigen-modes are significant. In order to
determine the significance of $\lambda_2,\lambda_3,\ldots$ and
$\vek{x}^{(2)},\vek{x}^{(3)},\ldots$, we generate random bipartite
graphs for comparison with the real data in the following way.
\begin{enumerate}
\item Cut every edge connecting bank $\mu$ and firm $i$. Then, for
  each original edge, we have
  two stubs; one from the bank (bank-stub) and the other from the firm
  (firm-stub).
\item Retain the original weight $w_{\mu i}$ on the $k_\mu$ stubs 
  emanating from the bank $\mu$.
\item Randomly choose a pair of a bank-stub and a firm-stub, and
  rewire the pair by an edge.
\end{enumerate}
The 3rd procedure is done so that there is no multiple edge between any
pair of bank and firm. This rewiring procedure alters the weight
as $w_{\mu i}\rightarrow w_{\mu j}$ if the edge emanating from $\mu$
to $i$ is randomly connected to $j$. Note that $w_\mu$, $k_\mu$ and
$k_i$ are invariant for each $\mu$ and $i$ under rewiring, while $w_i$
becomes randomized. Therefore, the matrix $\mat{A}$ has the same
structure except a permutation of columns. This means that a same
amount of credit is supplied by a bank to a different firm in the randomly
generated graphs.

The sum of eigenvalues satisfies \eref{eq:lamsum}. To compare the
spectrum $\lambda$ with that for random graphs, one has to do so
after a normalization. Define a normalized eigenvalue by
\begin{equation}
  \tilde{\lambda}_k=\frac{\lambda_k}{\displaystyle
    \sum_{\ell=1}^n \lambda_\ell}
\label{eq:lamnorm}
\end{equation}

\begin{figure}[!tb]
  \centering
  \includegraphics[width=0.80\textwidth]{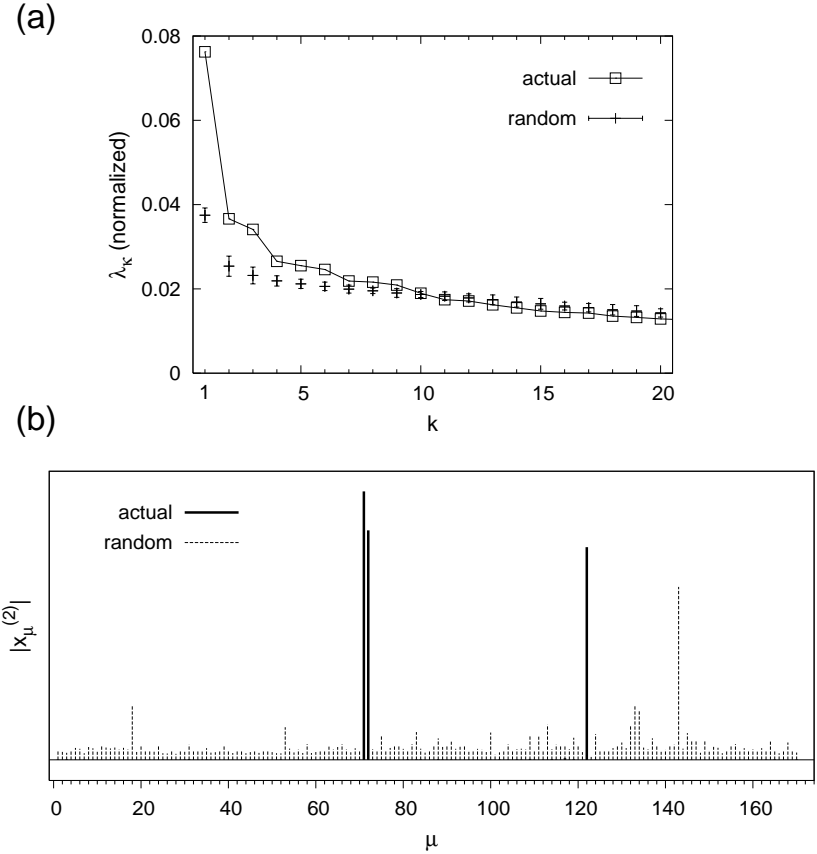}
  \caption{%
    (a) Largest 20 eigenvalues $\tilde{\lambda}_k$ defined by
    \eref{eq:lamnorm}. Squares are for the credit network in the year
    2005. The points are averages each for 10 realizations of random
    bipartite graphs with the standard deviation. %
    (b)~The components of eigenvector $|{x}^{(2)}_\mu|$ for the
    actual data in the year 2005 (solid lines). Dotted lines show
    absolute values of components averaged over the random graphs.}
\label{fig:lam_b-vec_b}
\end{figure}

\figref{fig:lam_b-vec_b}~(a) depicts the spectrum obtained for the
credit network in the year 2005. By comparing with the spectrum for
random graphs, we can say that only a few eigenvalues are significant.
In this case, they are $\tilde{\lambda}_2$ and $\tilde{\lambda}_3$
(except $\tilde{\lambda}_1=(\sum_\ell\lambda_\ell)^{-1}$), while
$\tilde{\lambda}_7$ and subsequent ones are indistinguishable from the
spectrum for random graphs.

The corresponding eigenvectors $\vek{x}^{(2)},\vek{x}^{(3)},\ldots$
have components at a set of banks. To show this, the components
$|{x}^{(2)}_\mu|$ is depicted in \figref{fig:lam_b-vec_b}~(b).
There are a few peaks at particular banks, while the same plot for
random graphs (absolute value of each component averaged over 10 randomly generated
graphs) is completely different from it.

This also demonstrates that these peaks of $|{x}^{(2)}_\mu|$ do
not simply reflect the distribution for $w_\mu$, because under the
randomization of bipartite graphs the configuration $w_\mu$ is not
altered at all.

We also remark that if one simply takes into account of connectivity
throwing away the information of weights, the resulting eigenvectors
have quite different characteristics. This can be readily verified by
assuming that $w_\mu=k_\mu$ and $w_i=k_i$, that is, by supposing that 
$w_{\mu i}=1$ for each edge.

\begin{figure}[!tb]
  \centering
  \includegraphics[width=0.70\textwidth]{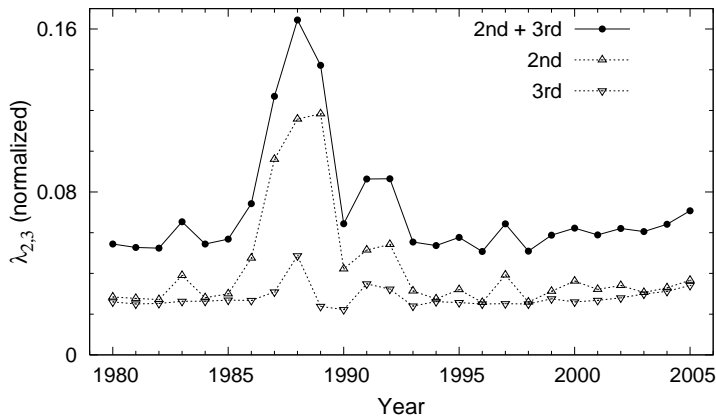}
  \caption{%
    The sum (solid line) of the normalized eigenvalues
    $\tilde{\lambda}_2$ and $\tilde{\lambda}_3$, with their values
    (triangles and dotted lines) in each year from 1980 to 2005.}
\label{fig:ts_lam23}
\end{figure}

For the historical data from 1980 to 2005, we obtained the spectrum in
each year to see how the eigenvalues change in time. The result is
shown in \figref{fig:ts_lam23} for the largest two eigenvalues
$\lambda_2$ and $\lambda_3$ normalized by \eref{eq:lamnorm}. There are
a strong peak in the late 80s and a drop in 1990; also two peaks
around 1992 and in 1997. 

The relationship between banks and firms changed in the course of the
Japanese bubble (speculative investment into stocks and real estate) in the
nation, notably in the late 80s up to 1990 and after the bubble (after
1990) period. Two points should be considered for understanding what
happened in the Japanese credit market during the period. First, firms
were allowed to issue public debt, after financial deregulation,
meaning that they were less dependent on bank loans. Secondly, after
the collapse of bubble, banks were left with non-performing loans,
which hindered the intermediary role of banks.  The problem of bad
loans affected individual firm's decision to contract banks. It is
known, for example, that during the bubble period the firms,
especially large ones, tended to rely
on a single relation, while in the period of long stagnation after the
collapse of the bubble the average percentage of multiple contracts
increases \citep{ogawa2007jfp}. Banks then spent a decade or longer in
90s to recover from bad loans experiencing a financial crisis for a
couple of years from 1997. In the late of 1997 and in 1998, three
major and other small banks failed. While two major banks were
nationalized, and other political decisions were made for maintaining
the stability of financial system, most banks, major and minor,
decreased the supply of credit immediately; even by reducing existing
loans to firms, most notably for small and medium-sized firms.

Our observation in \figref{fig:ts_lam23}, for the late 80s and an
abrupt change in 1990, coincides with this historical change of the
bank-firm relationship. The fragility score in terms of the
non-trivial eigenvalues increased during the period when firms tended
to have single relation. Also, in 1997, banks decreased the total
amount of loan during the 90s in attempt to reduce bad loans
systematically. This can be considered to decrease the diversity in
the credit system, resulting in the increase of fragility
score.

\begin{figure}[!tb]
  \centering
  \includegraphics[width=0.96\textwidth]{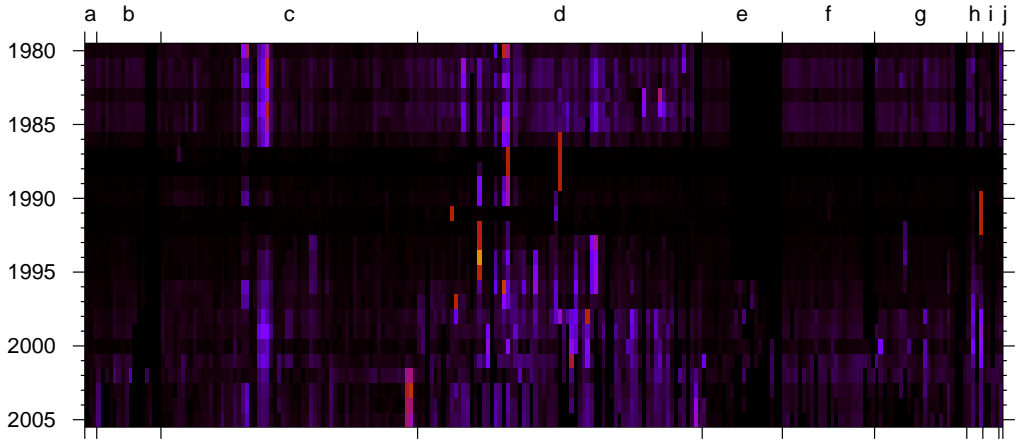}
  \caption{%
    Components of the non-trivial eigenvectors, $\vek{x}^{(2)}$ and
    $\vek{x}^{(3)}$, corresponding to the largest two eigenvalues,
    during the years from 1980 (top) to 2005 (bottom). Each row
    represents the average of $|{x}^{(2)}_\mu|$ and
    $|{x}^{(3)}_\mu|$ by color, while columns are for
    $\mu=1,\ldots,n$ ($n=229$). See \tabref{tab:banks} for the
    classification of financial institutions, {\sl a\/} (left) to {\sl
      j\/} (right). A cell's brighter color depicts a larger value.}
\label{fig:bcode-eig}
\end{figure}

We also examined the components of eigenvectors, $\vek{x}^{(2)}$ and
$\vek{x}^{(3)}$, in order to have a look at how stable or unstable the
eigen-structure is during the same period of time. We take the average
to have the information on how large the non-trivial eigenvalues are, in
comparison with that for random graphs, which can measure the fragility of
the credit network.  \figref{fig:bcode-eig} shows the average of
$|{x}^{(2)}_\mu|$ and $|{x}^{(3)}_\mu|$ for all the existed banks
$\mu$ (horizontally) in the years from 1980 to 2005 (vertically from
top to bottom). We can observe stable and unstable periods, and also
peaks at particular banks. Notably, unstable pattern can be observed
in the late 80s coinciding the course of the bubble, and also in 90s
after the bubble. There are peaks during these periods as well as for
1997 and 1998, which overlap the financial crisis. We shall
discuss more about the results in the next section.

\section{Discussion}\label{sec:disc}

\figref{fig:ts_lam23} and \figref{fig:bcode-eig} describe the temporal
change of the Japanese credit network with respect to the eigenvalues
and corresponding eigenvectors. In order to fully understand our
proposed scores of fragility for banks, one needs to compare the
scores with the characteristics of financial conditions of banks,
which can directly measure the level of financial deterioration. We
shall investigate this point elsewhere. Here we relate the obtained
results with historical description on the Japanese banking system in
the past 25 years.

The absolute values of the eigenvectors, in \figref{fig:bcode-eig},
have a relatively stable profile among banks from 1980 to 1986 and
from 2000 to 2005. The profile has peaks at several banks, notably a
few regional banks (in the middle-north geographical region). In the
late 80s, the profile changes dramatically, and spikes are
present at two banks, from 1986 to 1989, which are in the middle-north
region and are known to have deteriorated financially during the period.
In the late 80s to 90s, the Bank of Japan (BOJ) altered monetary
policy tightening the policy most notably in 1990. After the bubble
collapse, during the 90s, the profile changed into another
configuration. A spike in the classification of {\sl h\/} refers to
the Credit associations (Shinkin banks). Then, in the latter half of
90s, the profile went back to the previous one but with more peaks at
other regional banks (especially at secondary regional banks). The
spikes from 2003 to 2005 correspond to three banks in Okinawa.

Though we need more investigation beyond the anecdotal evidence,
it is intriguing to note that several of the spikes in the profiles
correspond to failed banks and to banks that had been merged into
larger banks.

Also we note that the peaks and spikes mentioned above are present in
same geographical regions --- middle and north regions, and Okinawa.
One of the authors (Y.~F.)  with collaborators recently showed that
banks can be clustered into groups according to their patterns of
lending to firms \citep{masi2008ajc}. In fact, by defining the pattern
for bank $\mu$ by a vector $\vek{a}_\mu$ that is equal to a column
vector of the matrix $\mat{A}$:
\begin{equation}
  (\vek{a}_\mu)_i:=A_{\mu i}\ ,
\end{equation}
it is possible to define a similarity in the lending patterns for a
pair of banks $\mu$ and $\nu$, for example, by the inner product of
the corresponding vectors $\vek{a}_\mu$ and $\vek{a}_\nu$. Then one
can perform the clustering by standard methods including
multi-dimensional scaling and hierarchical clustering. Indeed,
\citet{masi2008ajc} showed the minimum spanning tree (MST) calculated by a
similarity measure ignoring the information of weight but considering
only the connectivity from banks to firms. The resulting MST
corresponds to clusters of co-financing relationships of banks, which
strongly reflect the geographical regions especially for the regional
banks. It would be interesting to investigate how the eigen-structure
is related to those clusters.

It is also remarked that, as described in \secref{sec:data}, we did not
include the firms that went into bankruptcy. It should be interesting
to include them in the credit network in order to evaluate the effect
to banks and to compare the evaluation with the structural change that
followed after the bankruptcy. It would be possible to model such
propagation based on our consideration in defining the scores.

\section{Conclusion}\label{sec:conc}

We studied the structure and its temporal change of Japanese credit
relationships between commercial banks and quoted firms for the 25 years
from 1980 to 2005. Each snapshot of the credit network is regarded as a
weighted bipartite graph, where each node is either a bank or a firm,
and an edge between a bank and a firm is defined to be present if
there is a credit relationship between them. The edge has a weight
that represents the amount of credit.

Suppose that a bank shrinks the amount of its supplied credit, a firm
as debtor would be influenced to a certain extent that might be
quantified by a matrix that can be calculated by the weight.
Similarly, if a firm fails, then its effect to a bank as debtor would
propagate to an extent that is measurable from the weight. To quantify
the propagation, we introduced a set of score named fragility and
its dual, and proved mathematical properties among them. The set of
scores can be obtained by solving an eigenvalue problem.

By comparing the eigen-structure with that obtained in random
bipartite graphs, which have same distributions for degrees of banks
and firms and for normalized weight of banks, we found that the
largest few (non-trivial) eigenvalues for the scores are significant.
We performed historical analysis for our datasets, and showed that
there are periods when the eigen-structure is stable or unstable, and
that a particular set of banks, mostly a few regional banks, have
large values of the fragility scores. Drastic change occurs in the
late 80s during the bubble and also at the epochs of financially
unstable periods including the financial crisis. Further investigation
might be necessary to relate our results based on complex network
analysis to the characteristic of banks, but we believe that our
approach is a potentially valuable quantification
of the structure and its temporal change of credit relationships.

\appendix
\renewcommand{\thesection}{Appendix \Alph{section}:}
\renewcommand{\theequation}{\Alph{section}.\arabic{equation}}

\clearpage
\section{Mathematical properties of the eigenvalue problem}\label{sec:math}
\setcounter{equation}{0}

As shown in \secref{sec:nw_frag}, the set $\vek{x}$ of
fragility scores of banks is the right eigenvector of the weight
matrix $\mat{P}$ as in \eref{eq:eigx}, and the set $\vek{u}$ of dual
scores of banks satisfy the left eigenvector of $\mat{P}$ as in
\eref{eq:eigu}. In this Appendix, we prove mathematical properties on
eigenvalues and eigenvectors.

Let us first show that the score $\vek{u}$ can be calculated directly
from the score $\vek{x}$. \eref{eq:eigx} is written explicitly in
components as
\begin{equation}
  \frac{1}{w_\mu}\sum_{i,\nu}\frac{1}{w_i}w_{\mu i}w_{\nu i}x_\nu
  =\lambda x_\mu\ ,
\end{equation}
which we rewrite as
\begin{equation}
  \sum_{i,\nu}\frac{1}{w_i}w_{\mu i}w_{\nu i}x_\nu
  =\lambda w_\mu x_\mu\ .
\label{eq:eigxc}
\end{equation}
On the other hand, \eref{eq:eigu} is
\begin{equation}
  \sum_\mu u_\mu \frac{1}{w_\mu}\sum_i\frac{1}{w_i}w_{\mu i}w_{\nu i}
  =\lambda u_\nu\ ,
\end{equation}
which, after exchanging $\mu\leftrightarrow\nu$, reads as
\begin{equation}
  \sum_{i,\nu}\frac{1}{w_i}w_{\mu i}w_{\nu i}\frac{u_\nu}{w_\nu}
  =\lambda u_\mu\ .
\label{eq:eiguc}
\end{equation}
By comparing \eref{eq:eigxc} and \eref{eq:eiguc}, we find that
they are equivalent under the identification:
\begin{equation}
  u_\mu \propto w_\mu x_\mu\ .
\label{eq:xu}
\end{equation}
This also proves that left-eigenvalues and the right-eigenvalues
have a same spectrum.

Let us consider two sets of eigenvalues and corresponding
eigenvectors,\\
$(\lambda^{(k)},\vek{u}^{(k)},\vek{x}^{(k)})$ and
$(\lambda^{(\ell)},\vek{u}^{(\ell)},\vek{x}^{(\ell)})$. We have
\begin{equation}
  \vek{u}^{(k)\T}\,\mat{P}\,\vek{x}^{(\ell)}
  =\lambda^{(k)}\,\vek{u}^{(k)\T}\cdot\vek{x}^{(\ell)}
  =\lambda^{(\ell)}\,\vek{u}^{(k)\T}\cdot\vek{x}^{(\ell)}\ .
\end{equation}
This means that
\begin{equation}
  0=\left(\lambda^{(k)}-\lambda^{(\ell)}\right)
    \vek{u}^{(k)\T}\cdot\vek{x}^{(\ell)}
  =\left(\lambda^{(k)}-\lambda^{(\ell)}\right)
    \sum_{\mu}u_\mu^{(k)} x_\mu^{(\ell)}\ ,
\end{equation}
which, by the use of \eref{eq:xu}, implies that
\begin{equation}
  0=\left(\lambda^{(k)}-\lambda^{(\ell)}\right)
    \sum_{\mu}w_\mu x_\mu^{(k)}x_\mu^{(\ell)}\ .
\end{equation}
Therefore, the eigenvectors should be orthonormal under the weight
$w_\mu$ as a metric%
\footnote{%
  Mathematically, $\vek{x}$ is a covariant vector, $\vek{u}$ is a
  contravariant vector, and the metric that connects them is given by 
  $g_{\mu\nu}=\delta_{\mu\nu}\,w_{\mu}$. The orthogonalization of
  eigenvectors is done with respect to this metric.}%
. That is,
\begin{equation}
  \sum_{\mu}w_\mu x_\mu^{(k)} x_\mu^{(\ell)}=\delta_{k\ell}\ .
\label{eq:ortx}
\end{equation}
It follows from \eref{eq:ortx} the orthonormality:
\begin{equation}
  \sum_{k}w_\mu x_\mu^{(k)} x_\nu^{(k)}=\delta_{\mu\nu}\ .
\label{eq:ortx2}
\end{equation}

This consideration of the inner product implies that we should take a
look at the product of \eref{eq:eigxc} and $x_\mu$. This leads us to
\begin{equation}
  \lambda=\frac{\ \displaystyle\sum_i\frac{1}{w_i}
    \biggl(\sum_{\mu}w_{\mu i}x_\mu\biggr)^2 \ }%
  {\displaystyle \sum_{\mu}w_\mu x_\mu^2}\ .
\label{eq:lamtmp}
\end{equation}
This proves that $\lambda$ is real and positive, although the matrix
$\mat{P}$ is not symmetric. Also we have the following inequality
that holds for any value of $q$.
\begin{equation}
  0\leq\sum_\mu w_{\mu i} (q-x_\mu)^2
  =w_i q^2-2\biggl(\sum_{\mu}w_{\mu i}x_\mu\biggr)q
   +\sum_\mu w_{\mu i} x_\mu^2\ .
\end{equation}
This leads to the inequality for the discriminant:
\begin{equation}
  \biggl(\sum_{\mu}w_{\mu i}x_\mu\biggr)^2
  -w_i\sum_\mu w_{\mu i} x_\mu^2 \le 0\ ,
\end{equation}
from which it proves that the largest eigenvalue is 1.
\begin{equation}
  0 < \lambda\leq 1\ .
\end{equation}
This proves \eref{eq:spectrum}. It is obvious from \eref{eq:lamtmp}
that $\lambda=1$ if and only if $x_\mu=q$. In fact, one can easily
see, from \eref{eq:sum_A} and \eref{eq:sum_B} that $x_\mu=1$
($\mu=1,\ldots,n$) is the eigenvector corresponding to $\lambda=1$,
provided that the bipartite graph is connected (i.e. any node of bank
or firm is reachable from any other)%
\footnote{%
  For a disconnected graph, $x_\mu$ is constant in each connected
  components. The multiplicity of $\lambda=1$ is equal to the number
  of the connected components.}%
. This proves \eref{eq:trivia}.

In addition, by applying the orthogonal relation in \eref{eq:ortx2} to
\eref{eq:eigxc}, it can be shown after a short calculation that the
summation formula holds:
\begin{equation}
  \sum_{k}\lambda_k=\sum_{\mu,i}A_{\mu i}B_{i\mu}
  =\textrm{tr}\,\mat{P}\ .
\label{eq:lamsum:app}
\end{equation}
This proves \eref{eq:lamsum}.

To summarize, the eigenvector $\vek{u}$ can be calculated directly
from the eigenvector $\vek{x}$. Also the eigenvalues satisfy
$0<\lambda\leq 1$, where the largest eigenvalue corresponds to a
trivial eigenvector.

On the other hand, the dual scores, $\vek{u}$, corresponding to the
largest eigenvalue $\lambda=1$ simply represents the total amount of
loans, namely $u_\mu\propto w_\mu$ due to \eref{eq:xu}, so we can focus on non-trivial
eigenvectors, $\vek{x}^{(2)}$, $\vek{x}^{(3)}$ and so on in the main
text.

\newpage
\ifx\undefined\bysame
\newcommand{\bysame}{\leavevmode\hbox to\leftmargin{\hrulefill\,\,}}
\fi


\end{document}